# Sky Brightness Measurements and Ways to Mitigate Light Pollution in Kirksville, Missouri

**Vayujeet Gokhale**
**Jordan Goins**
**Ashley Herdmann**
**Eric Hilker**
**Emily Wren**
*Truman State University, 100 E Normal Street, Kirksville, MO 63501; gokhale@truman.edu*

**David Caples**
**James Tompkins**
*Moberly Area Community College, Kirksville, MO 63501*

**Abstract** We describe the level of light pollution in and around Kirksville, Missouri, and at Anderson Mesa near Flagstaff, Arizona, by measuring the sky brightness using Unihedron sky quality meters. We report that, on average, the Anderson Mesa site is approximately 1.3 mag/arcsec$^2$ darker than the Truman State Observatory site, and approximately 2.5 mag/arcsec$^2$ darker than the roof of the science building at Truman State University in Kirksville. We also show that at the Truman observatory site, the North and East skies have significantly high sky brightness (by about 1 mag/arcsec$^2$) as compared to the South and West skies. Similarly, the sky brightness varies significantly with azimuth on the top of the science building at Truman State—the west direction being as much as 3 mag/arcsec$^2$ brighter than the south direction. The sky brightness at Anderson Mesa is much more uniform, varying by less than 0.4 mag/arcsec$^2$ at most along the azimuthal direction. Finally, we describe the steps we are taking in the Kirksville area to mitigate the nuisance of light pollution by installing fully shielded outdoor light fixtures and improved outdoor lights on Truman State University's campus.

## 1. Introduction

Light pollution is the introduction of artificial light, either directly or indirectly, into the natural environment. It refers to wasted light that performs no useful function or task and leads to light trespass, and increased sky glow and glare. While having some level of outdoor lighting is prudent for aesthetic reasons and for public safety, badly designed light fixtures are wasteful and lead to decrease in human, animal, and plant well-being. Most of the wasted light comes from outdoor lighting such as residential lights, street lights, and business lights. While the problem of light pollution is of particular importance to the astronomy community, the safety, health, ecological, and environmental costs of light pollution are gaining increasing attention (Gaston *et al.* 2012; Chepesiuk 2009). The problem of light pollution is expected to become even more acute with the growing use of increasingly affordable and efficient LED lighting, which will not only affect the quantity of light pollution (brightness), but also the quality of the light (color). In addition to disrupting the sleep cycles of humans, light pollution also disrupts ecological systems (Sanchez *et al.* 2017; Aube *et al.* 2013). Natural patterns of light determine wildlife behavior in a variety of ways, including predation, reproduction, and fatigue, which are drastically altered when subjected to light pollution. Furthermore, light pollution comes at the price of unnecessary energy costs and carbon emissions, impacting both the consumer and the environment (Gaston *et al.* 2012).

In the town of Kirksville, Missouri, the University itself is a major source of light pollution, and poses several problems to students living on or close to campus. Fiscally, this can be seen in the electricity usage of the University, which is incorporated into the prices of room and board, as well as tuition. Light trespass from unshielded light shines directly into the windows of residence halls, disrupting the sleep patterns of students. An additional problem inflicted by light pollution is glare, which can impair the vision of drivers passing through campus, putting pedestrians at risk. Commonly used outdoor lights emit a significant amount of light in wavelengths shorter than $\approx 500$ nm, toward the blue-end of the electromagnetic spectrum. Blue light scatters more than red light, and hence using outdoor lights which emit more energy towards the blue-end (wavelength $< 500$ nm) of the electromagnetic spectrum causes greater sky glow than lights emitting most of their energy towards the red end (wavelength $\geq 500$ nm). Additionally, it is known that blue light causes more glare than red light (Intl. Dark Sky Assoc. 2010) and so it is prudent to use outdoor lights which emit less light towards the blue-end of the electromagnetic spectrum. Thus, using lights which emit most of their energy at wavelengths greater than 500 nm effectively reduces the sky brightness and is comparatively soothing to the eye (Intl. Dark Sky Assoc. 2010; Luginbuhl *et al.* 2010; Mace *et al.* 2001).

There are several effective methods of light pollution reduction. These include retrofitting existing fixtures with fully shielded light shields which direct light towards the ground, and installing outdoor lights with color temperature $T \leq 3000$ K, with lesser emission at wavelengths < 500 nm than the outdoor lights currently in use. As mentioned, a significant contribution of outdoor lighting is from street and business lights, and any change in the quality and quantity of outdoor lighting will have to involve the active involvement of local, regional, and federal government authorities. One way to garner public support, and to convince authorities to make the necessary changes in policies, is providing evidence for light pollution via long-term monitoring of the sky brightness at several locations. This will ensure mitigation of any biases introduced by a particular location (proximity to playground lights, for example) as well as due to temporal factors (moon phase, cloud cover, decorative lights during holidays, and so on). With this in mind, we have devised a three-point plan to quantify and mitigate the nuisance of light pollution in the Kirksville (Missouri, USA) community that involves:

1. Quantifying light pollution in and around the town of Kirksville, Missouri, using Unihedron SQMs,
2. Increasing awareness about the nuisance and dangers of light pollution, and
3. Working with city and school authorities to transition to outdoor lights with color temperature less than 3000 K, and install light shields and light friendly fixtures to reduce light pollution.

In this paper, we describe the current light pollution level in and around Kirksville, Missouri (population approximately 17,000), and at Anderson Mesa, about 15 miles southeast of Flagstaff, Arizona (population approximately 72,000). Kirksville presents a semi-rural setting in the north of the state of Missouri, while Flagstaff is a designated dark-sky location with light ordinances and zoning codes. We compare the night sky brightness levels in Kirksville with similar measurements made near Flagstaff by using Unihedron (http://unihedron.com/index.php) sky quality meters (SQMs) installed at various locations. Unihedron manufactures different types of SQM sensors though, for practical reasons, we favor the handheld version for Alt-Az measurements and the datalogging SQM-LU-DL for continuous sky brightness measurements. We describe some of the properties of the SQMs we use and our set-up in the following section. In section 3, we present our results and analyses of SQM measurements made at various sites over the past few years. In section 4 we discuss our ongoing efforts and future plans regarding the quantification and mitigation of light pollution. In particular, we describe how we have involved students and student organizations and used them as leverage to push administrators towards installing light shields and improved outdoor lighting on our campus and downtown area.

## 2. Sensor properties and set-up

For this project, our main concern was to ensure that the sensors are mutually consistent, so that we could compare the sky brightness at various locations on a given night. In order to test this, we periodically set up the sensors right next to each other and compared the SQM readings from these sensors. In all we have used five sensors named *Yoda, Darth, Rey, Finn,* and *Kylo*. All of these are datalogging SQMs (SQM-LU-DL) which allow for continuous monitoring of the sky. These sensors have a full-width-at-half-maximum of about 20°. The SQM sky brightness is given in units of magnitudes per square arcsecond (mags/arcsec$^2$), which means that a difference in 5 units is equivalent to a ratio of 100 in luminance (for details, please see Kyba *et al.* (2011).

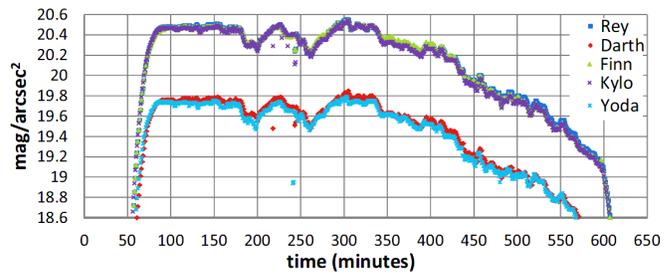

Figure 1. A typical plot showing the sky quality measurements using five sensors placed at the same location (Truman State Observatory). The newer sensors (*Rey, Finn, Kylo*) give darker measurements than do the older sensors *Yoda* and *Darth*. See text for details.

From the simultaneous runs of these sensors at the same location, we noticed that the two older sensors, *Darth* and *Yoda*, were consistently giving us values between 0.65 to 0.75 "higher" than the newer sensors (*Rey, Finn,* and *Kylo*) when the sky brightness is about 19 mags/arcsec$^2$ or darker (Figure 1). The variations between the three new sensors are minimal—at most 0.1 unit when the sky brightness is about 19 mags/arcsec$^2$ or darker. Consequently, for the sake of consistency, we added an offset of 0.7 mag/arcsec$^2$ to all our readings obtained from *Darth* and *Yoda* while comparing sky brightness at various locations. Note that we found that the offset is not constant at different levels of darkness and may have a temperature-dependence as well. (We have informed Unihedron about this discrepancy. Tekatch (2019) informs us that the most common issue is of darker readings caused by a frosted IR sensor, though it is unclear if this is the issue with our sensors. We intend to take up Unihedron's offer to recalibrate and clean the optics (free of charge) in order to identify and correct the problem.) For future studies, we intend to discard the older sensors and closely monitor and calibrate the sensors to ensure they give consistent results to avoid this problem.

Data were collected at numerous sites in two cities - Kirksville, Missouri, and Flagstaff, Arizona. Kirksville is a small town (population ≈ 17,000, elevation 300 m) in northeast Missouri, while Flagstaff is a "dark sky city" in northern Arizona (population ≈ 72,000, elevation 2,130 m). The different sites (see Table 1 and Figure 2) in Kirksville included the roof of the science building (Magruder Hall, MG from here on), the roof of the authors' residence (VG-roof) located 5 miles northwest of campus, and the Truman State Observatory (TSO), located about 2 miles south west of campus. The MG site is surrounded on all sides with unshielded "globe" lights (Figures 2 and 10) which are a significant source of light pollution. The Anderson Mesa (AM) site, the location of Lowell Observatory research telescopes, is about 12 miles southeast of Flagstaff and is considered a rural "dark site."

Table 1. Geographical characteristics of sites used in this work.

| Site | Nearest City | Geographic Coordinates (°) | | Elevation | Comments |
|---|---|---|---|---|---|
| AM | Flagstaff, Arizona (North-East) | 35.0553 N | 111.4404 W | 2163 m, 7096 ft | dark site, usually low humidity |
| MG | Kirksville, Missouri | 40.1866 N | 92.5809 W | 299 m, 981 ft | urban site, usually high humidity |
| TSO | Kirksville, Missouri (North-East) | 40.177 N | 92.6010 W | 299 m, 981 ft | semi-rural site, usually high humidity |
| VG Roof | Kirksville, Missouri (South-East) | 40.2110 N | 92.6305 W | 299 m, 981 ft | semi-rural site, usually high humidity |

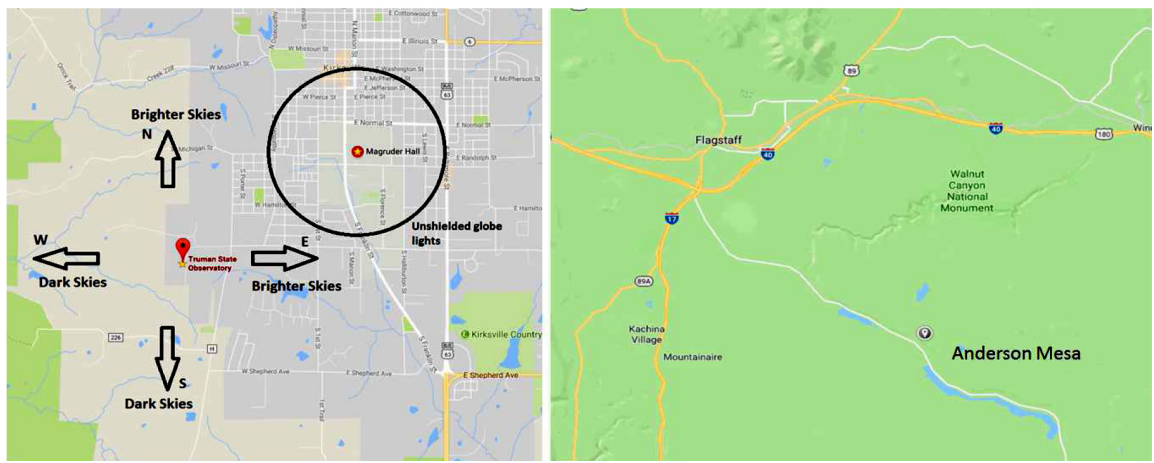

Figure 2. Geography of the Kirksville sites (left panel) and the Anderson Mesa site near Flagstaff, Arizona (right panel).

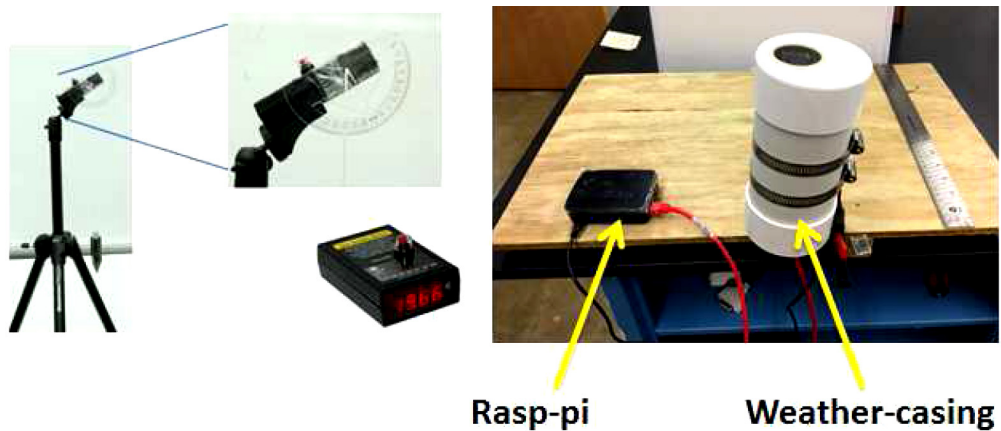

Figure 3. Left Panel: Set up for sky brightness measurements as a function of altitude-azimuth using hand-held SQMs Birriel and Adkins (2010). Right Panel: SQM-LE sensor inside the weather casing attached to a raspberry-pi via an Ethernet cable for continuous measurements. The SQM-LU-DL USB-connected sensor has a similar setup.

## 3. Quantifying sky brightness

In this paper, we describe two ways in which we quantified the sky brightness. In the first method, we use the manually operated Unihedron SQM light sensor (Half Width Half Maximum (HWHM) of the angular sensitivity is ≈ 42°) and the SQM-L sensor (HWHM of the angular sensitivity is ≈ 10°) to measure the sky brightness as a function of the altitude in four directions (east, west, north, and south). (Two different versions of the manually operated SQM were used since in 2017 we only had access to the SQM. The SQM-L sensor became available to us only after 2018.) In the second method, we use the "automatic" datalogging SQM-LU-DL sensor housed in a weather-proof case to monitor the sky brightness. These datalogging sensors can be powered by batteries which allow us to measure sky brightness at remote locations over several nights.

### 3.1. Angle dependence

We measured the sky brightness as a function of the altitude and azimuthal angle following the procedure outlined by Birriel and Adkins (2010). The manually operated SQM is mounted on a tripod with a clear protractor attached horizontally to it. A plumb bob is suspended using a string to enable accurate measurements of the zenith angle (see Figure 3). As much as possible, measurements were made on moonless, cloudless nights away from trees, tall walls, and buildings. At each angle, we recorded the sky brightness five times, and calculated the average before moving on to a different angle.

The angles were changed by either 10 or 15 degrees. We carried out these measurements at three different locations: the roof of the science building (MG hall), the Truman State Observatory, and at Anderson Mesa near Flagstaff, Arizona. The differences in the three sites are striking (Figures 4, 5, and 6). As shown in Figure 2, the science building is surrounded on all sides by dark-sky unfriendly "globe" lights and consequently the sky brightness is significant at low altitude (this is partly due to reflection from buildings and trees, see Figure 4). The TSO location has high sky brightness in the direction of Kirksville, which lies to its north-east (Figure 2). The sky brightness does not appear to depend strongly on altitude toward the south and west due to lack of any street lights or residences in those directions (Figure 5). The data from the two epochs are largely consistent, except close to the horizon where local land features (trees, bushes, street lights) might affect the measurements. At the TSO site (Figure 5), there is considerable discrepancy in the "east" data from the two epochs, of the order of 1 mag/arcsec$^2$ near the horizon. This can be attributed to street lights on a road that runs north-south to the east of the TSO site. Also, note that the 2017 data were collected using the SQM meter which has a much greater angular sensitivity than the SQM-L meter used for obtaining the 2019 data, which explains the "brighter" readings on the SQM meter when pointed in the direction of the town, though a similar effect is not seen in the north direction.

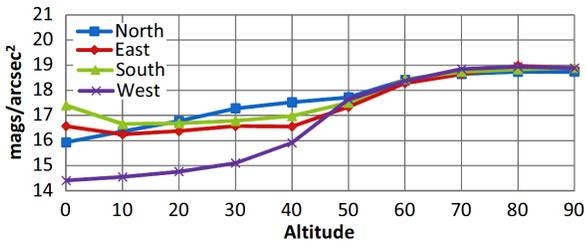

Figure 4. Typical plot of the measurements of sky brightness as a function of altitude and direction at the MG site using the SQM sensor. The sky is darkest close to the zenith, and brightest along the horizon. The sky brightness variation in a given direction is between 4.5 mag/arcsec$^2$ (west) and 1.5 mag/arcsec$^2$ (south). The sky brightness is the greatest toward the west due to the presence of a large number of unshielded "globe" lights in that direction.

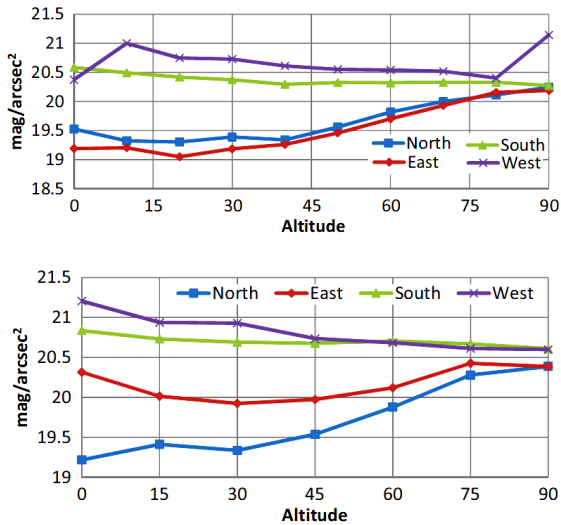

Figure 5. Sky brightness measurements as a function of altitude and direction at the TSO site from data collected approximately two years apart using the SQM (upper panel) and SQM-L (lower panel) sensors. Note the high sky brightness levels in the east and north, the direction of Kirksville town. The presence of a state park and wilderness towards the west and south directions results in relatively darker skies in these directions

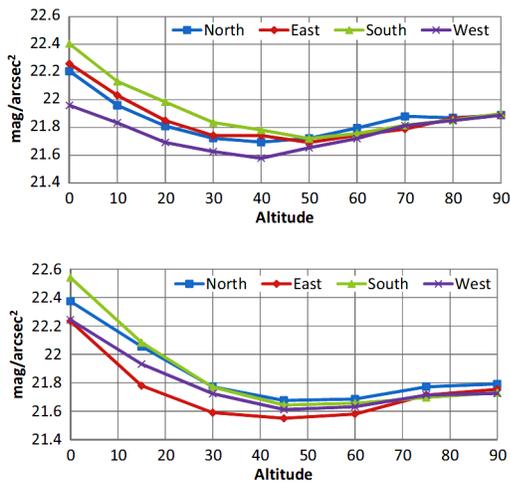

Figure 6. Same as Figure 5, but at the AM site near Flagstaff, Arizona on two clear night two years apart. Note the relatively short range of variation in the sky brightness as compared to the MG location. On average, the sky brightness levels at AM are about 3 mag/arcsec$^2$ better at the zenith, and as much as 8 mag/arcsec$^2$ better at the horizon than the MG location (see Figure 4).

The altitudinal dependence of sky brightness in Flagstaff is somewhat different than the city and semi-rural locations in Kirksville. At the Anderson Mesa location near Flagstaff, absence of street lights in all directions results in a significantly dark horizon, with gradually increasing sky brightness to about 45° in altitude. This increase in brightness may be a result of natural or artificial sky glow, or from near or far field luminance and illumination (Schaefer 2019; Walker 1977). The sky brightness then decreases toward the zenith. The AM-site is south-east of Flagstaff, resulting in slightly poorer skies towards the north and west (the Moon was rising in the east during the 2019 measurements, biasing our results in that direction).

Our results qualitatively match that of Birriel and Adkins (2010), though there are quantitative differences due to the geography vis-a-vis terrain and the location of town with respect to the measurement site. In particular, the variation in the sky brightness in the azimuthal direction is much greater closer to town (≈ 2 mag/arcsec$^2$, their Figure 3a) as against at a location farther from town (≤ 1 mag/arcsec$^2$, their Figure 2a). This is comparable to the results presented here in Figures 4, 5, and 6: at the most light polluted location, MG roof, the azimuthal variation is the largest at the horizon (≈ 2.7 mag/arcsec$^2$), at the semi-rural location (TSO) the azimuthal variation is about 2 mag/arcsec$^2$, while at the rural, "dark sky" location, the azimuthal variation is ≤ 0.5 mag/arcsec$^2$. In terms of altitudinal variation, Birriel and Adkins (2010) do not observe a brightening at around 45° at the "dark sky" location (their Figure 3) as we do at the AM-site (Figure 6). This could be because the AM-site is much darker (21.9 mag/arcsec$^2$ at zenith) than the Cave Run Lake (21.2 mag/arcsec$^2$ at zenith) site.

3.2. Continuous monitoring

The second method we are using to measure sky brightness is by using the "datalogging" mode of the SQM-LU-DL sensor provided by Unihedron. This sensor can be powered by a battery and is capable of storing data for several nights. Encased in the weather-proof casing, this set up is ideal for sky-brightness measurements in remote locations without power and/or wireless internet. We use this sensor in two ways: one is battery operated, and in the other mode, we connect the SQM-LU-DL to a Raspberry-pi microcomputer, which then transmits data via wireless to a webserver (Figure 3). We set up the SQM-LU-DL in its weather-proof case, pointing toward the zenith, at a location away from any trees, buildings, and overhead lights. Within city limits, the sensor is always mounted on the roof of a convenient, accessible building above the level of the street lights, while in semi-rural and rural areas the sensor is mounted on a laboratory-stand placed on an even surface, usually the ground.

Figures 7 and 8 show plots of some of our "continuous monitoring" runs. These data were taken over the past few years at various locations in varying weather conditions. Results from the AM-site are fairly consistent over the measurements made two years apart. The sky brightness is about 21.9 mag/arcsec$^2$ in the absence of clouds and the moon. On the other hand, the data from the different locations in Kirksville show a significant level of light pollution. The TSO location, about two miles south west of campus, is the darkest site, with a sky brightness of about 20.7 mag/arcsec$^2$, while the VG-roof location is about 20.4 mag/arcsec$^2$. The MG-site, located on the Truman State University campus, has a sky brightness of about 19.4 mag/arcsec$^2$, approximately 2.5 units worse than the AM-site. Again, the data are fairly consistent several months apart, which is reassuring.

Note that passing clouds are evident in the plots shown below (Figures 7, 8, and 9)—light is reflected back toward the Earth from the bottom of the clouds, resulting in an uneven sky brightness curve as measured by the SQMs (Kyba *et al.* 2011). In general, the sky brightness increases as a consequence of clouds—especially so in light polluted areas since there is more ambient light to reflect back from the bottom of the clouds. This is evident in 2019 data shown in Figure 8. On this night, clouds cleared up by timestamp 300 leading to a leveling out of the SQM readings at the three different sites around Kirksville. Note, however, that at the AM-site, the sky brightness actually decreased (right panel, Figure 7) on 27 May 2019, a significantly cloudy night. This may be due to the lack of light projected upward due to the absence of outdoor light sources near the AM-site. It is also possible that the sky brightness is affected differently by different kinds of clouds (high cirrus, stratus, cumulus, etc.), something that we are investigating presently.

Figure 9 shows a comparison of the sky-brightness measured at various locations in Kirksville and at Anderson Mesa near Flagstaff. It is clear that the AM-sky is darker by about 1.3 mags/arcsec$^2$ as compared to the semi-rural sites in Kirksville. Also, the AM-site is, on average, about 2.5 mags/arcsec$^2$ darker than MG-hall on the Truman State campus.

### 4. Discussion and future work

We are currently in the process of installing the SQM-LU-DL sensors at various locations across the town of Kirksville, Missouri. We intend to have three permanent SQM-LU-DL sensors attached to Raspberry-pi micro-computers that can automatically transmit data via wireless internet to a webserver, two of which are currently operational at the TSO and MG-hall sites. These long-term measurements can serve as a baseline for comparison in the sky brightness levels at different locations, as well as for comparison between measurements made several years apart at the same location. In turn, these can then be used to convince authorities to improve outdoor lighting and to provide objective evidence for the effectiveness of using fully shielded light fixtures, once they are implemented.

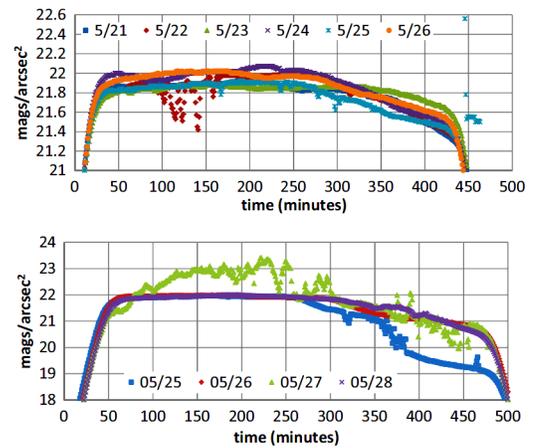

Figure 7. SQM-LU-DL measurements at Anderson Mesa for six nights in May 2017 (upper panel) and four nights in May 2019 (lower panel) as a function of time. Some of nights were cloudy, leading to several "spikes" in the sky brightness measurements. For 2017, moonrise is around timestamp 300 (2:40 AM MST) on 21 May (dark blue curve), while moonrise is around timestamp 250 for 25 May 2019.

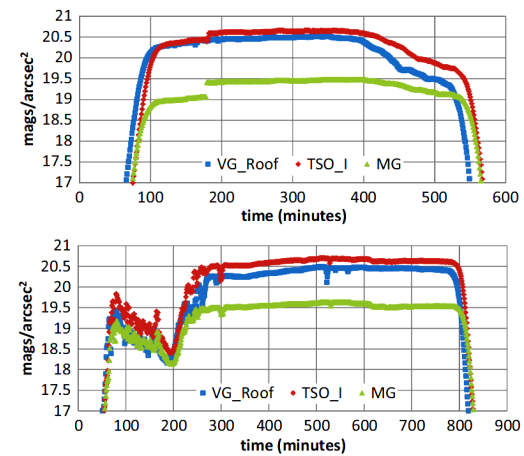

Figure 8. Different SQM-LU-DL measurements in Kirksville at four different locations on 8 May 2018 and 7 January 2019. The urban location (MG) has the greatest sky brightness, while the semi-rural location at the Truman Observatory (red) is comparably dark. The light blue curve is the sky brightness from the roof of the author's residence, at a location 5 miles west/north-west of Kirksville. On the upper panel, an abrupt jump can be seen in the sky brightness around timestamp 175 at locations close to the Truman campus (most prominent in the TSO-curve, but can also be seen in the MG-curve) corresponding to the switching "off" of the lights on the football stadium on campus. The lower panel shows the effects of clouds at the beginning of the night, until approximately timestamp 300 after which the sky was reasonably clear and the SQM curves flatten out.

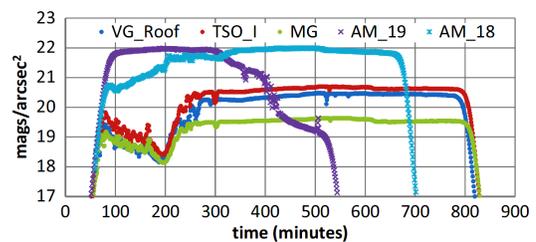

Figure 9. Comparison between various locations using the datalogging SQM-LU-DL sensors. The Anderson Mesa site is clearly the darkest of the sites we monitored. The green dots represent data from the roof of Magruder Hall, the science building on the Truman State University campus, which is by far the most light-polluted site. The data were staggered to align the onset of darkness after sunset—the different durations of darkness represent the differing duration of night. The Truman data were taken in early January, while the Anderson Mesa data are from March 2018 and May 2019.

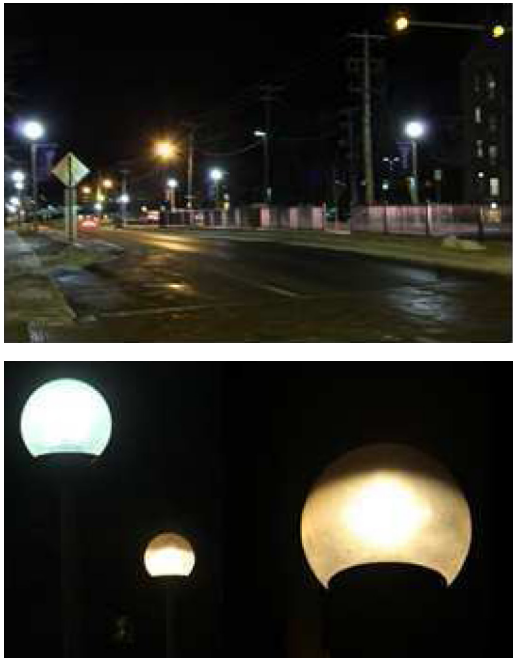

Figure 10. Upper Panel: Unshielded "globe" lights galore on Franklin Street. One of the student dorm buildings can be seen toward the right of the image. Lower Panel: An example of an unshielded light fixture ("globe lights") housing a $T$ = 5000 K (blue/white color) light bulb and another fixture fitted with a "dark sky reflector" housing a $T$ = 3000 K (off-white) light bulb. $T \leq 3000$ K light is known to cause less glare and is less detrimental to the environment and plant and animal health (Intl. Dark Sky Assoc. 2010; Luginbuhl *et al.* 2010).

We have recently set up a weather station, cloud sensor, and an all-sky-camera at the Truman State Observatory. We are currently analyzing data from the SQMs installed at the Observatory and correlating these data with weather parameters such as humidity, temperature, and cloud cover. The all-sky-camera images provide a visual check to the cloud cover measurements made by the cloud sensor. Preliminary results show a strong dependence of the sky brightness on cloud cover (Kyba *et al.* 2011}. We are particularly interested in monitoring the effects of different kinds of clouds (cirrus, stratus, cumulus, etc.) on sky brightness.

In addition to the quantitative and qualitative elements of the research, we have devoted a significant amount of time and effort toward increasing awareness about light pollution. This has involved engaging and educating both the Kirksville and Truman populations about the harmful effects of light pollution. Students from the Truman State "Light Pollution Group" make presentations about the detrimental effects of light pollution to supplement shows at the Del and Norma Robison Planetarium at Truman State University. IDA brochures (https://www.darksky.org/our-work/grassroots-advocacy/resources/public-outreach-materials/) are distributed to the audience after a viewing of the IDA documentary "Losing the Dark" (https://www.darksky.org/our-work/grassroots-advocacy/resources/losing-the-dark/).

We are in the process of acquiring night-sky friendly light shields (Figure 10) to cover some of the unshielded light fixtures on campus. In collaboration with the "Stargazers" student astronomy group on campus, we obtained funding for these light shields from the Environmental Sustainability Fee Committee (EFC) and the Funds Allotment Council (FAC) at Truman State University. Both these funds are generated via a nominal (≈ $5 per semester) fee imposed on each student attending Truman State. The proposals for these grants were written by participating students. The level of funding from the EFC and the FAC is $6,000 and $2,500, respectively. These funds were used to purchase IDA-approved dark-sky shields to retrofit the "globe" lights on campus to reduce skyglow. Starting fall 2019, fifty such shields are being installed in a selected area (Franklin Street, Figure 10) to test the effectiveness of the shields in terms of student approval, reduction of skyglow, and maintenance factors. Franklin Street was chosen due to the presence of several unshielded lights close to on-campus housing halls. If the light-shields are successful in combating light pollution and glare, we plan on applying for additional funds to purchase and install more shields. In addition, we are transitioning to improved outdoor lighting (The lights being installed (color-temperature $T \approx 3000$ K) emit about 15% of their light at wavelengths below 500 nm (Maa 2019) by replacing the blue/white lamps currently in use on most outdoor light fixtures. These actions present us with an opportunity to do before-and after-studies to investigate the impact of the dark-sky reflectors and the changed lighting on the sky brightness.

As outlined in the Introduction (section 1) we have made significant progress in our three-step program to address the issue of wayward outdoor lighting. We have successfully set up several SQM meters at various locations to quantify sky brightness and have engaged in creating awareness. We have used our data and analyses to convince authorities to implement dark-sky friendly light fixtures. In collaboration with the Missouri chapter of the International Dark Sky Association (https://darkskymissouri.org/), we are working on establishing a network of SQM sensors across several cities, parks, and recreational areas across the state of Missouri in the near future. We hope that in the next few years, we can establish several "dark-sky friendly" parks and recreational areas in the state of Missouri by significantly reducing light pollution. Though it exists world-wide, light pollution remains a local problem with local solutions. While the data presented in this paper are specific to measurements made at the particular locations under consideration, we believe that the overall rationale, methodology, and analyses presented here can be duplicated at other locations. We hope that concerned citizens, and amateur and professional astronomers across the nation and the world will follow suit and work towards creating safe, night-sky friendly environments in their communities. Human beings have evolved and grown up with an unadulterated view of the beautiful night sky for millennia. We owe it to ourselves, and to future generations, to be not deprived of this beauty.

## 5. Acknowledgements

This work is supported by funds from the Missouri Space Grant Consortium and the Provost's office at Truman State University. The authors are thankful to the support from the Environmental Sustainability Fee Accountability Committee and the Funds Allotment Council at Truman State University for funding to purchase the dark sky reflector light-shields and new $T$ = 3000 K outdoor lights. VG would like to thank the Physical Plant personnel at Truman State University for their willingness to work with us on this project. The authors would also like to thank Andrew Neugarten for setting up the Raspberry-pi microcomputers. The authors would also like to thank the anonymous referee for useful comments and suggestions, which greatly improved the manuscript.

**References**

Aube, M., Roby, J., and Kocifaj, M. 2013, *PLoS ONE*, **8**, e67798.

Birriel, J., and Adkins, J. K. 2010, *J. Amer. Assoc. Var. Star Obs.*, **38**, 221.

Chepesiuk, R. 2009, *Environ. Health Prospect.*, **117**, A20.

Gaston, K. J., *et al.*. 2012, *J. Appl. Ecology*, **49**, 1256.

International Dark Sky Association. 2010, "Visibility, Environmental, and Astronomical Issues Associated with Blue-Rich White Outdoor Lighting" (https://www.darksky.org/our-work/grassroots-advocacy/resources/ida-publications/).

Kyba C. C. M., Ruhtz, T., Fischer, J., and Holker, F. 2011, *PLoS ONE*, **6**, e17307.

Luginbuhl, C., Moore, C., and McGovern, T., eds. 2010, *Nightscape*, No. 80, 8 (https://www.darksky.org/wp-content/uploads/bsk-pdf-manager/3_SEEINGBLUE.PDF).

Maa, M. 2019, private communication.

Mace, D., Garvey, P., Porter, R. J., Schwab, R., and Adrian, W. 2001, "Countermeasures for Reducing the Effects of Headlight Glare" (https://trid.trb.org/view/707950), AAA Foundation for Traffic Safety, Washington, DC.

Sanchez de Miguel, A., Aubé, M., Zamorano, J., Kocifaj, M., Roby, J., and Tapia, C. 2017, *Mon. Not. Roy. Astron. Soc.*, **467**, 2966.

Schaefer, B. 2019, private communication.

Tekatch, A. 2019, private communication.

Walker, M. F., 1977, *Publ. Astron. Soc. Pacific*, **89**, 405.